\documentclass[
   ,final
  ]
  {aipproc}

\layoutstyle{6x9}

\begin{document}

\title 
      {Particle production at RHIC energies}

\author{R. Debbe for the BRAHMS collaboration}{
   address={Physics Dept. Brookhaven National Laboratory}
} 

\copyrightyear  {2003}

\begin{abstract}

This paper presents recent results from the BRAHMS experiment at RHIC; including results on particle 
production in rapidity space extending from y=0 to $y\sim 3$ and on the
transverse momentum distribution of fully identified charged particles. These results were 
obtained from the $5\%$ most central Au-Au collisions recorded
during RHIC Run-2 at $\sqrt{s_{NN}} = 200$ GeV.

\end{abstract}

\maketitle

\section{Introduction}

BRAHMS is the only RHIC experiment that is able to study fully identified particle production and 
energy flow over a wide range of rapidity (from y=0
to y=4 for pions). This coverage, which  almost reaches the fragmentation 
regions, is ideal for studies of the bulk properties of the system formed in 
heavy ion collisions at RHIC energies. 
This work reports recent results obtained from the analysis of data collected 
with the BRAHMS spectrometers in
Au-Au collisions at $\sqrt{s_{NN}} = 200$ GeV. A detailed description of the 
BRAHMS experimental setup can be found in \cite{NIM}. All results shown here
are preliminary and were obtained from a sample of the $5\%$ most central 
events.
\vspace{-0.5cm}    

\section{Particle production}

Momentum distributions of fully identified charged particles were obtained with
 conventional magnetic spectrometers instrumented with  state-of-the-art 
time-of-flight 
and ring imaging and threshold \v{C}erenkov 
detectors.  For each particle type, the density in rapidity space is 
obtained by
integration over the  ${p}_{\rm{T}} $ dependence of these distributions. With the 
distributions dropping rapidly as  ${p}_{\rm{T}} $ increases, a good fraction of the integral 
comes from unmeasured yields at low  ${p}_{\rm{T}} $ . An extrapolation is thus necessary to cover the
unmeasured regions. Empirically the
pion yields were found to be best fitted by a power law, the kaon distributions by a single exponential in $p_T$, and the proton distributions with a single exponential in $m_T$.

 The resulting rapidity density distributions are shown in panels a and b of
figure 1. Panel a shows the densities for all charged particles. Because 
pions dominate this figure, panel b expands the view for the kaon, proton and antiproton
distributions.

\begin{figure}
\resizebox{0.7\textwidth}{!}
          {\includegraphics{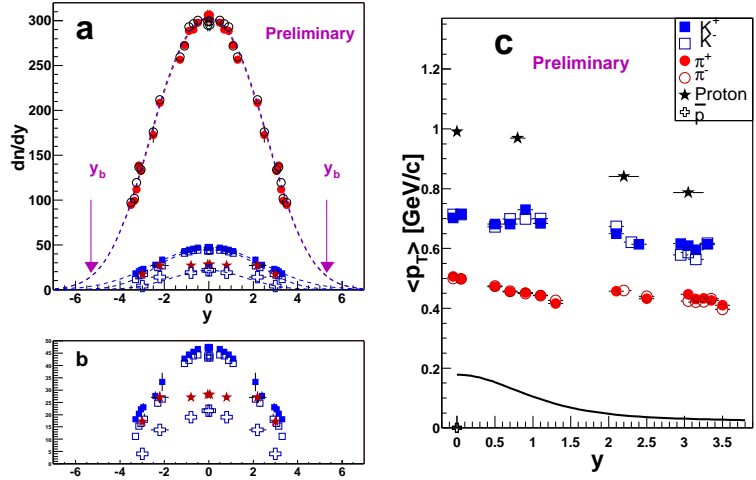}}
\caption{a) and b)Symmetrized rapidity density distributions for identified charged particles measured in the $5\%$ most central events. The measurements were done for y>0. c) Mean ransverse momentum as
function of rapidity}
\end{figure}

 A remarkable feature of these distributions is their common bell shape character. (Pions, kaons and anti-proton  distributions are fitted well with double Gaussians).
 This observation has a  possible explanation based on the postulate that
particle production in the  momentum range measured in the present 
experiment is driven primarily by the distribution of partons in the colliding
ions. 
 As the energy of the collisions increases, the parton distributions can be
resolved to smaller values of x (fraction of the total 
momentum of the hadron). The convolution of the left and right moving parton distributions leads to an initial bell-shaped dn/dy  distribution for produced
particles in symmetric system centered around y=0.

This distribution may evolve in later stages through secondary interactions, 
but
it retains its bell shape. In this picture there is neither a wide plateau connecting the two 
fragmentation regions, as Feynman's intuition had it \cite{Feynman}, nor an 
extended boost
invariant longitudinal expansion, as proposed by Bjorken \cite{Bjorken}.

A good summary of all the $p_T$ distributions extracted in this analysis is
shown in panel c of figure 1; the average transverse momentum with which
the detected particles are produced at different rapidities. Worth noting in
this result is the small change of the pion and kaon
average $p_{\rm{T}}$ as function of y.
For comparison, a calculated average $p_T$ for pions is also drawn in panel c. The curve
was obtained with a single thermal source described by a Boltzmann distribution
with a temperature of 200 MeV.

An inspection of the transverse momentum distributions shows that 
for the more massive particles there is a very clear curvature
in the spectra as the value of  ${p}_{\rm{T}} $ approaches zero (see Figure 2).
 This observation, together with
the almost exponential shape of the distributions at higher ${p}_{\rm{T}} $ and for lower mass particles, is well reproduced by a 
functional form based on a thermalized system expanding 
radially \cite{HeinzFlow}.

After integration over y and azimuthal angle and assuming that a Boltzmann 
distribution describes the system, the following 
functional form is obtained, and is used to fit the spectra:

$$ \frac {dn}{m_T dm_T} = A m_T \int_0^R rdr K_{1}(\frac{m_{T}cosh\rho}{T}) I_{0}(\frac{p_{T}sinh\rho}{T}) $$   

where $ \beta_T = tanh \rho$ is the transverse velocity of the flow in 
units of c, and T is the decoupling temperature.

\begin{figure}
\resizebox{0.7\textwidth}{!}
          {\includegraphics{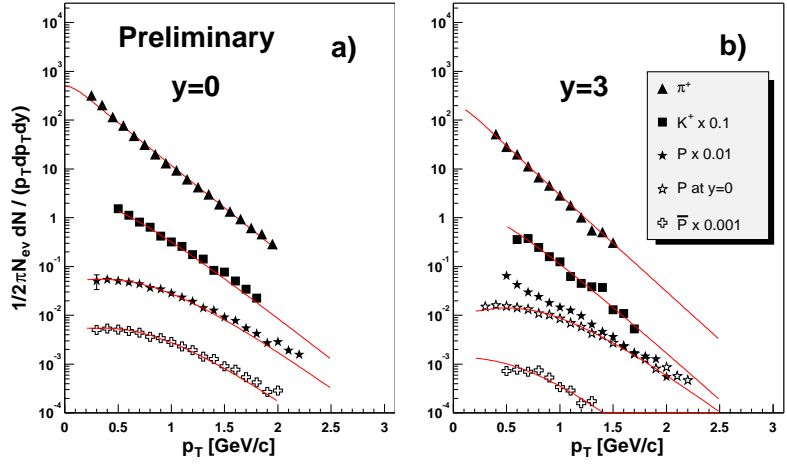}}
\caption{``Blast wave'' fits to  pions, kaons and protons at rapidity y=0 
(panel a) and y=3 (panel b)}
\end{figure} 

Several functions that describe the radial dependence of the flow velocity have
been proposed. Here the simplest assumption of a transverse flow 
velocity that is constant at all radii is assumed. This choice may not be 
fully adequate, but does serve to highlight the rapidity dependence of the flow velocity.
The table 1 summarizes the results.

\begin{table}
\begin{tabular}{lrr}
\hline
\tablehead{1}{r}{b}{Rapidity}
&\tablehead{1}{r}{b}{Temperature [MeV]}
&\tablehead{1}{r}{b}{Velocity} \\
\hline
0 &$ 127\pm2 $& $0.57\pm0.01$ \\
0.7 & $112\pm1$ & $0.60\pm0.01$ \\
2.2 & $128\pm 3 $& $0.50\pm 0.01$ \\
3 &$ 136 \pm 4$ & $0.44 \pm 0.02$ \\
\hline
\end{tabular}
\caption{Results of blast wave fits}
\end{table}
The shapes of the distributions are well reproduced by the fits, with a
 30 \% reduction in transverse velocity found going from y-0 to y=3. This 
reduction is also suggested by the different shape of the mid-rapidity proton 
distributions as 
compared to that at the most forward rapidity, as shown in Fig. 2b where the y=0 points have been shifted down to facilitate the comparison.

The measured net proton at mid-rapidity \cite{STAR},\cite{PHENIX} was an early subject of much discussion
in the community because it went against an expected baryon free region 
around $y \sim 0$; the energy of the colliding beams was so high that the 
initial baryon number should end up in the 
fragmentation regions if tied to the valence quarks . But the first results extracted at y=0 indicated 
otherwise, the net-proton number was not equal to zero. Some mechanism was 
transporting baryon number to mid-rapidity.
D. Kharzeev \cite{Kharzeev} had predicted that the x distributions
of  gluons of the initial
 baryon extend to very small values of x. At the time of the collision these
distributions would 
overlap and after ``dressing'' with quarks from the sea the low-x gluons of the initial baryons bring in effect 
net baryon number to mid-rapidity. This effect would have a rapidity dependence
as $\frac{Z b}{sinh(y_{max}b)}cosh(yb) $ where Z is the charge of the ions, 
and b is related to nature of the system that brings baryon number to mid-rapidity and predicted to be close to 1/2. The parameter $y_{max}$ is set to be equal to 4.5. Figure 3 shows the net charge measured up to 
rapidity 3 and a fit to the function mentioned above, The fit parameters are
 $b =  0.49 \pm 0.10 $  with $\chi^2 = 1./7$.

\begin{figure}
\resizebox{0.7\textwidth}{!}
          {\includegraphics{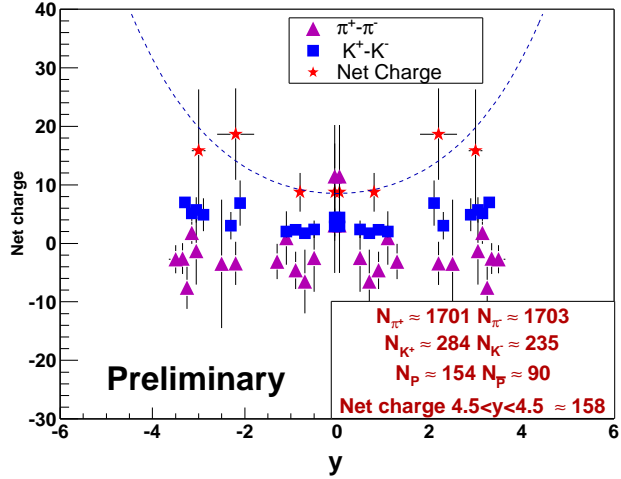}}
\caption{Net charge as function of rapidity ( star symbols). The 
difference $K^+ - K^-$ is shown with squares, and the 
triangles show the difference $\pi^+ - \pi^-$. The dashed line is the fit to the
function refered in the text. }
\end{figure} 

\vspace{-0.6cm}
\section{Summary }

Analysis of the most central data sample collected during RHIC Run-2 (Au-Au at
 $\sqrt{s_{NN}} = 200$ GeV) is consistent with a thermalized system with a 
decoupling temperature around 120 MeV, and a strong radial flow 
$\beta_{\rm{T}} \sim 0.6$ that diminishes by as much as 30\% at the most 
forward rapidity measured.

\vspace{-0.6cm}

\begin{theacknowledgments}
This work was supported by the Division of Nuclear Physics of the Office of Science of the U.S. 
DOE, the Danish Natural Science Research Council, the Research Council of Norway,
the Polish State Com. for Scientific Research, and the Romanian Ministry of
Education and Research.
\end{theacknowledgments}

\end{document}